\documentclass[conference]{IEEEtran}

\ifCLASSOPTIONcompsoc
 
  \usepackage[nocompress]{cite}
\else
  \usepackage{cite}
\fi

\usepackage{graphicx}
\usepackage{xcolor}
\usepackage{amsmath}
\usepackage{relsize}
\usepackage{array}
\usepackage{scalerel}
\usepackage{tikz}
\usetikzlibrary{svg.path}
\definecolor{orcidlogocol}{HTML}{A6CE39}
\tikzset{
  orcidlogo/.pic={
    \fill[orcidlogocol] svg{M256,128c0,70.7-57.3,128-128,128C57.3,256,0,198.7,0,128C0,57.3,57.3,0,128,0C198.7,0,256,57.3,256,128z};
    \fill[white] svg{M86.3,186.2H70.9V79.1h15.4v48.4V186.2z}
                 svg{M108.9,79.1h41.6c39.6,0,57,28.3,57,53.6c0,27.5-21.5,53.6-56.8,53.6h-41.8V79.1z M124.3,172.4h24.5c34.9,0,42.9-26.5,42.9-39.7c0-21.5-13.7-39.7-43.7-39.7h-23.7V172.4z}
                 svg{M88.7,56.8c0,5.5-4.5,10.1-10.1,10.1c-5.6,0-10.1-4.6-10.1-10.1c0-5.6,4.5-10.1,10.1-10.1C84.2,46.7,88.7,51.3,88.7,56.8z};
  }
}

\newcommand\orcidicon[1]{\href{https://orcid.org/#1}{\mbox{\scalerel*{
\begin{tikzpicture}[yscale=-1,transform shape]
\pic{orcidlogo};
\end{tikzpicture}
}{|}}}}
\usepackage{hyperref}
\usepackage{url}
\ifCLASSINFOpdf
\usepackage{threeparttable}
\usepackage{fancyhdr,lipsum}
\fancypagestyle{firstpage}{
\fancyhf{}

\fancyhead[C]{Accepted as conference paper at the IEEE International Symposium on Circuits and Systems (ISCAS) 2020}
\fancyfoot[C]{\footnotesize{\copyright 2020 IEEE. Personal use of this material is permitted. Permission from IEEE must be obtained for all other uses, in any current or future media, including reprinting/republishing this material for advertising or promotional purposes, creating new collective works, for
resale or redistribution to servers or lists, or reuse of any copyrighted component of this work in other works.
}}
}
\begin{document}

\title{Unified Characterization Platform for Emerging NVM Technology: Neural Network Application Benchmarking Using off-the-shelf NVM Chips}

\author{\IEEEauthorblockN{Supriya~Chakraborty \orcidicon{0000-0001-8890-5223}\,, Abhishek~Gupta, and ~Manan~Suri \orcidicon{0000-0003-1417-3570}\,}
\IEEEauthorblockA{Department of Electrical Engineering, Indian Institute of Technology Delhi, New Delhi, India\\ Email: manansuri@ee.iitd.ac.in}}

\maketitle
\thispagestyle{firstpage}


\begin{abstract}
In this paper, we present a unified FPGA based electrical test-bench for characterizing different emerging Non-Volatile Memory (NVM) chips. In particular, we present  detailed electrical characterization and benchmarking of multiple commercially available, off-the-shelf, NVM chips viz.: MRAM, FeRAM, CBRAM, and ReRAM. We investigate important NVM parameters such as: (i) current consumption patterns, (ii) endurance, and (iii) error characterization. The proposed FPGA based testbench is then utilized for a Proof-of-Concept (PoC) Neural Network (NN) image classification application. Four emerging NVM chips are benchmarked against standard SRAM and Flash technology for the AI application as active weight memory during inference mode. 
\end{abstract}

\section{Introduction}
The increasing trend of memory content in System-on-Chip (SoC) designs demand embedded or off-the-shelf Non-Volatile Memory (NVM) with low power consumption, high speed operations, and high endurance \cite{marinissen2005challenges,meena2014overview}. Flash EEPROM is the current state-of-art NVM technology used for commercial and industrial SoCs. However, Flash memories suffer from limitations such as physical scaling, erase-before-write operation, limited endurance, cell to cell interference, high power consumption, low programming speed, complex controller structures. The limitations of the Flash memories are overcome by exploring emerging NVM technologies such as magnetoresistive random access memory (MRAM), resistive RAM (ReRAM), ferroelectric RAM (FeRAM), conductive bridge RAM (CBRAM) \cite{hong2010memory,li2013nand,lu2009future,chen2016review}. 

Most of the research work on emerging NVM technologies in literature are based on model simulations of device and circuit, at architectural/system level \cite{kang2015modeling,padilha2018structure,dong2012nvsim,xie2011modeling}, single standalone device or an array of devices \cite{pellizzer200690nm,sheu20095ns}. Researchers are focused on improving the emerging NVM technologies at different levels like materials \cite{zhu2015overview},  stack engineering \cite{chand2017enhancement}, and circuit-level strategies \cite{mao2016optimizing}. Moreover, testing of matured NVM technology requires efficient characterization setup. Single standalone setup for characterizing multiple NVM technologies is rare to find. Commercially available packaged memory chip testing setups are complex and dedicated to a particular memory technology \cite{siglead,cai2011fpga,bunker2012ming}. The data-sheet specifications of the commercially available NVM chips provide typical and maximum values for write current and endurance of the NVM chips. However, variations may occur when the chips are used for real-time applications. The major factors include i) incoming data to be written in the memory location and ii) aging effect. Detailed characterization of the NVM chips is required for system-level integration in a variety of applications.  In this paper, we present a unified test platform for characterizing multiple commercially available off-the-shelf NVM technologies. Our study helps in exploring electrical and endurance properties of fabricated NVM chip and exploits them by designing hardware/software techniques for performance enhancement. The proposed hardware setup can be used generically for characterizing off-the-shelf emerging NVM chips (SPI or parallel interface). Moreover, the setup presents an efficient way to indirectly extract certain analog characteristics from the packaged chip without having access to a dedicated analog interface. The contributions of this paper are

\begin{figure}
  \centering
  \includegraphics[scale=0.25]{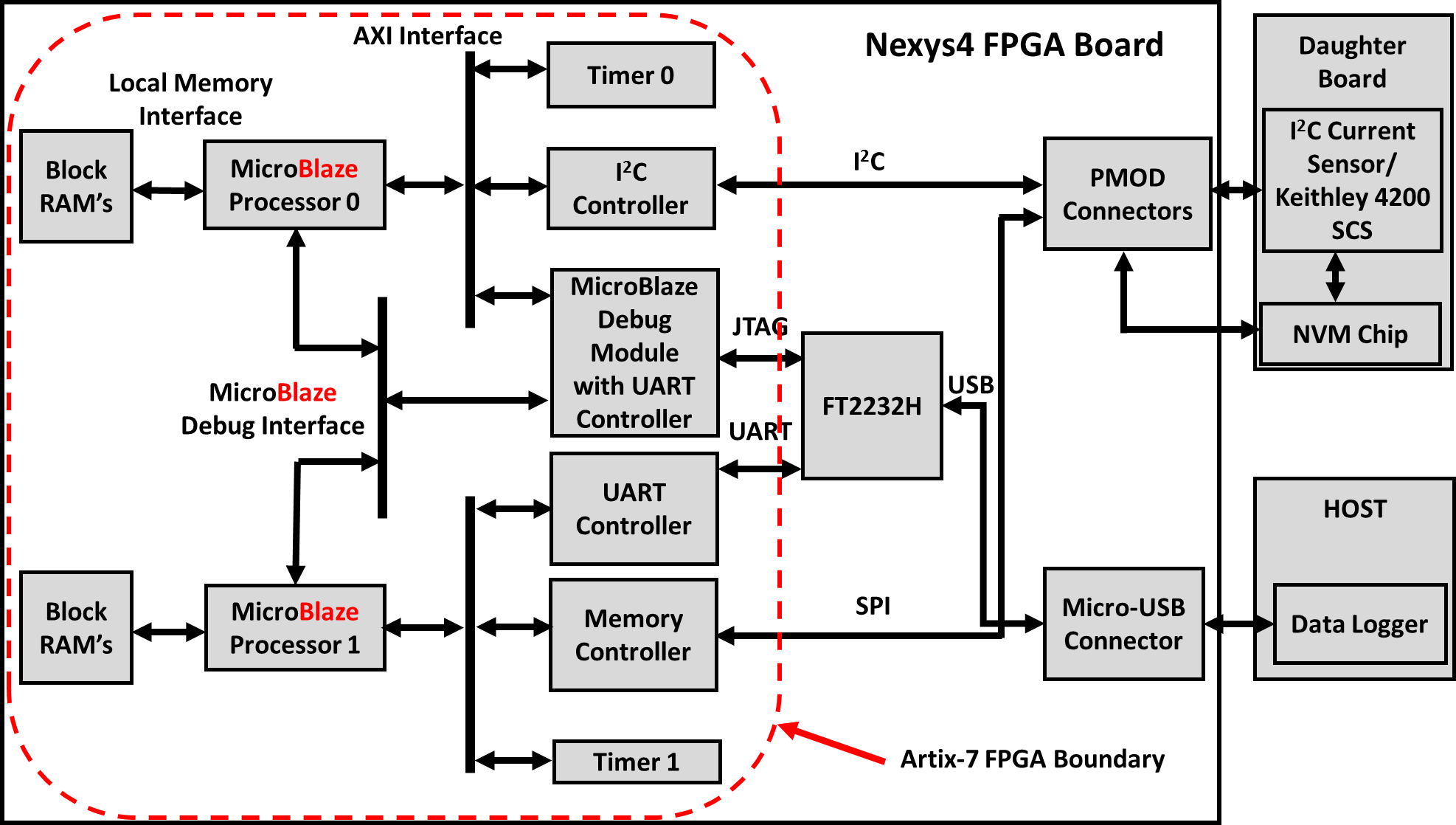}
    \caption{Block diagram of our experimental setup for NVM characterization.}
    \label{set_up}
        \vspace{-5mm}
\end{figure}
\begin{enumerate}
\item Electrical characterization of  different commercially available emerging NVM technologies specifically (1) toggle MRAM \cite{everspin}, (2) FeRAM \cite{cypress}, (3) CBRAM \cite{adesto}, and (4) ReRAM \cite{fujitsu} from different vendors. 
\item Error characterization of NVM chips.
\item Implementation of the proposed FPGA based setup for basic neural network (NN)  application case study comprising of a hybrid CMOS-NVM pipeline.   
\end{enumerate}
The rest of the paper is organized as follows: Section II explains the experiments performed on the emerging NVM technologies. Results and discussions are presented in Section III. A case study of NVM technologies for basic NN application using the proposed characterization platform is presented in Section IV. Section V concludes the paper. 
\section{Experiment Performed}
\subsection{Characterization Platform}
The characterization platform (shown in Fig. \ref{set_up}) comprises of FPGA evaluation board, custom designed daughter board to interface the NVM ICs, Keithley 4200 SCS characterization system for current measurement and a host computer. Two soft-core MicroBlaze processors are implemented on the FPGA to control digital interfaces between the host computer, NVM chips and other peripheral ICs. Both the soft-core processors with dedicated hardware are used to run parallel tasks. One of the processors (Processor 1) performs continuous read/write/erase operations from/to the NVM chips. Another processor (Processor 0) reads the write current measured using the current sensor IC. The operating clock frequency of the NVM chips (SPI based) is set to 1.5625 MHz for all the experiments. Write current is measured using two different methodologies: (i) using on-board I$^{2}$C interface based current sensor \cite{current_sensor} to measure the write current of toggle MRAM.
(ii) using Keithley 4200 SCS instrument to measure write current of FeRAM, CBRAM, and ReRAM. One terminal of the current characterization instrument is connected to supply voltage from the FPGA board while the other terminal of the instrument is connected to the VDD supply pin of the NVM chips. The host machine collects the data using virtual serial port connection. A detailed description of similar kind of experimental setup and procedure for current measurement and endurance characterization is explained in \cite{chakraborty2018current,suri2017experimental}. 

\subsection{Current Characterization}
Write current variations in NVM technologies due to input data pattern is analyzed by performing extensive data write operations with fixed number of bits toggled per byte. Initially, all the bits in a byte are set to either `0s' or `1s' followed by toggling specific bit(s) in a byte. Write current is measured for a particular page (64 bytes). The write operations are performed for multiple times (500 cycles) to measure the average write current for a particular type of bits toggled. The same experiment is repeated for different data patterns varying the number of bits toggled (from 1 to 8) in a byte. Write current variation due to the aging of the memory devices is performed by writing random data at a particular location for multiple (more than 50k) cycles.

\subsection{Endurance Characterization}
Error characterization is analyzed by performing extensive data write operations at a particular location (address) in the chip.
For each experiment, random data write operations are performed in a page for 200k cycles. A read operation is performed between two consecutive write cycles. Error is estimated by calculating the number of bits found incorrect between the data that is ``to be programmed" and the data that is ``actually programmed" at any specific byte address. We characterize and term the nature of the error based on the number of bits found incorrect in a byte. For example, in a byte, if 3-bits are incorrect we termed it as a 3-bit error. 

\begin{figure}[!t]
  \centering
\includegraphics[scale=0.33]{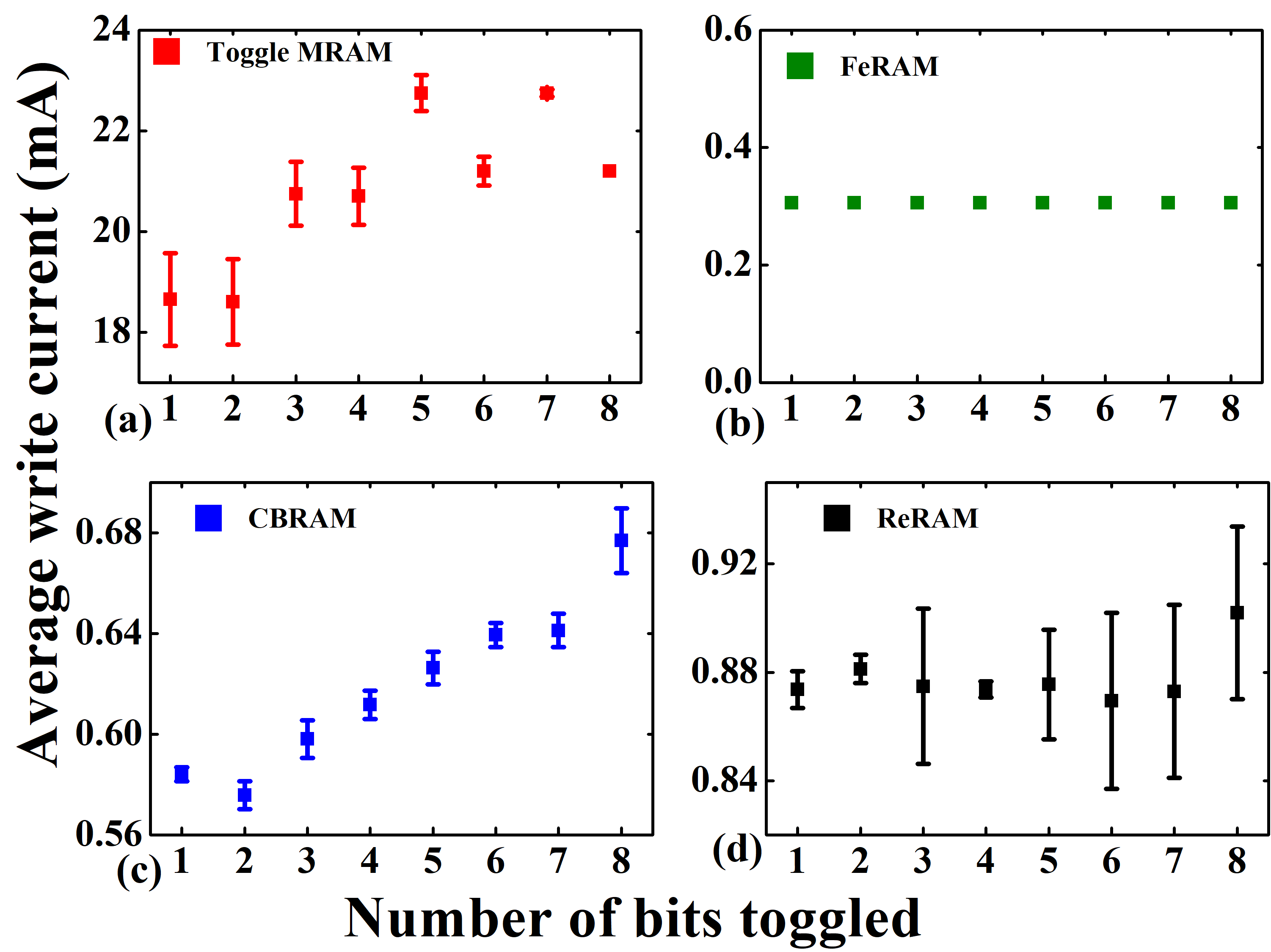}
    \caption{Variation of average page write current with increasing number of bits toggled for different NVM chips.}
    \label{write_current}  
\end{figure}
\begin{figure}[!t]
  \centering
 \includegraphics[scale=0.24]{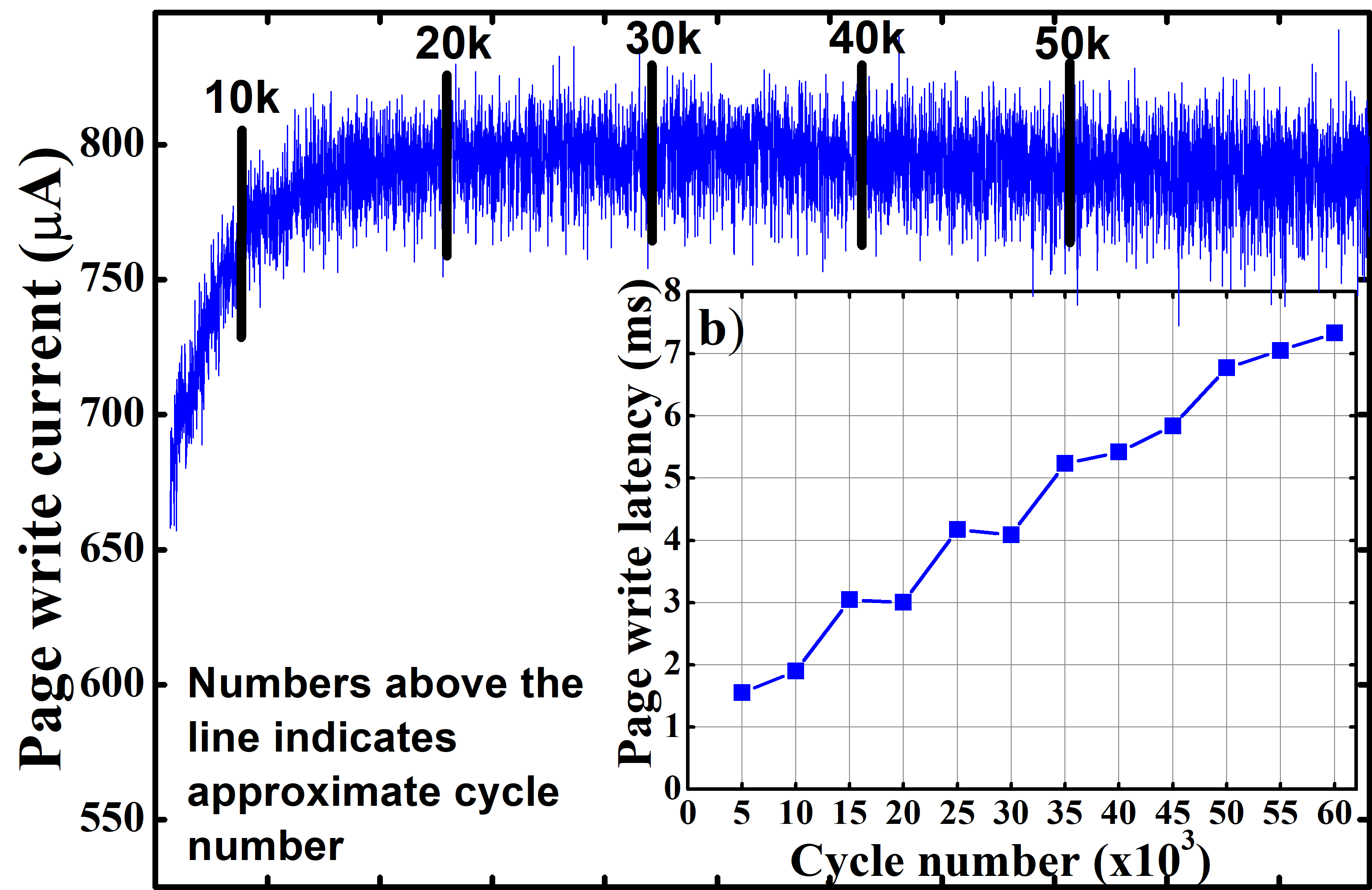}
    \caption{(a) Variation of page write current consumption with cycling in CBRAM. (b) Increasing pattern of write latency with cycling.}
    \label{aging_current_cbram}    
    \vspace{-5mm}
\end{figure}

\section{Results and Discussion} 
\subsection{Current Variation with Data Pattern }
Average page write current for different emerging NVM technologies with increasing number of bits toggled is shown in Fig. \ref{write_current}(a)-(d). It can be observed that page write current variation depending upon the number of the bits toggled exists and is different for different NVM technologies. 
Fig. \ref{write_current}(a), shows the current variation for toggle MRAM with increasing number of bits toggled. It can be observed that current deviation decreases with increase in number of bits toggled. This can be explained with the fact that in toggle MRAM, current consumption of toggled bit(s) with data pattern having more number of 1’s in the initial conditions is more compared to data pattern having more number of 0’s in the initial condition. For example, toggling of 1-bit for data pattern 11111111 consumes more current as compared to toggling of 1-bit for data pattern 00000000. Fig. \ref{write_current}(c) shows that page write current increases linearly with increase in number of bit toggled. This deterministic pattern as observed for the write current in toggle MRAM and CBRAM can be exploited to save power during data write operations. This can be obtained by implementing encoding data write operations based on least bits toggled \cite{chakraborty2018current}. No significant signature is observed for current consumption in writing increasing order toggled bit data pattern in FeRAM (Fig. \ref{write_current}(b)). The current consumption for writing any data pattern is constant and does not vary with number of bits toggled. However, observed magnitude of the page write current consumption is low thus can be used for low power applications. Page write current consumption for ReRAM is random in nature for writing different bits toggled as shown in Fig. \ref{write_current}(d).

\begin{figure}[!t]
  \centering
  \includegraphics[scale=0.35]{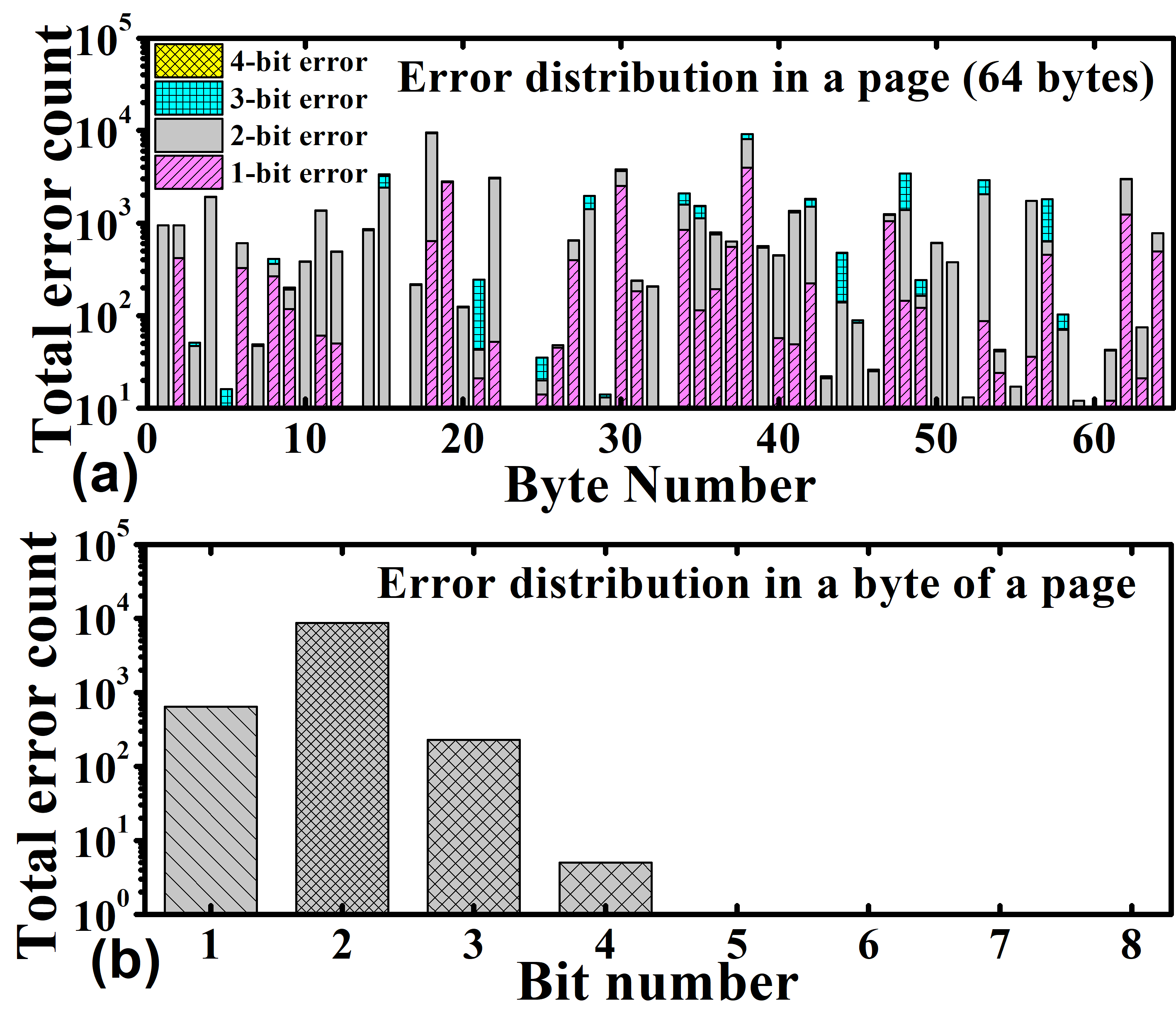}
    \caption{Nature of error distribution of (a) a  page (64 bytes), and (b) a byte program for 200k cycles in CBRAM.}
    \label{error_colour_page}
\end{figure}

\begin{figure}[!t]
  \centering
  \includegraphics[scale=0.25]{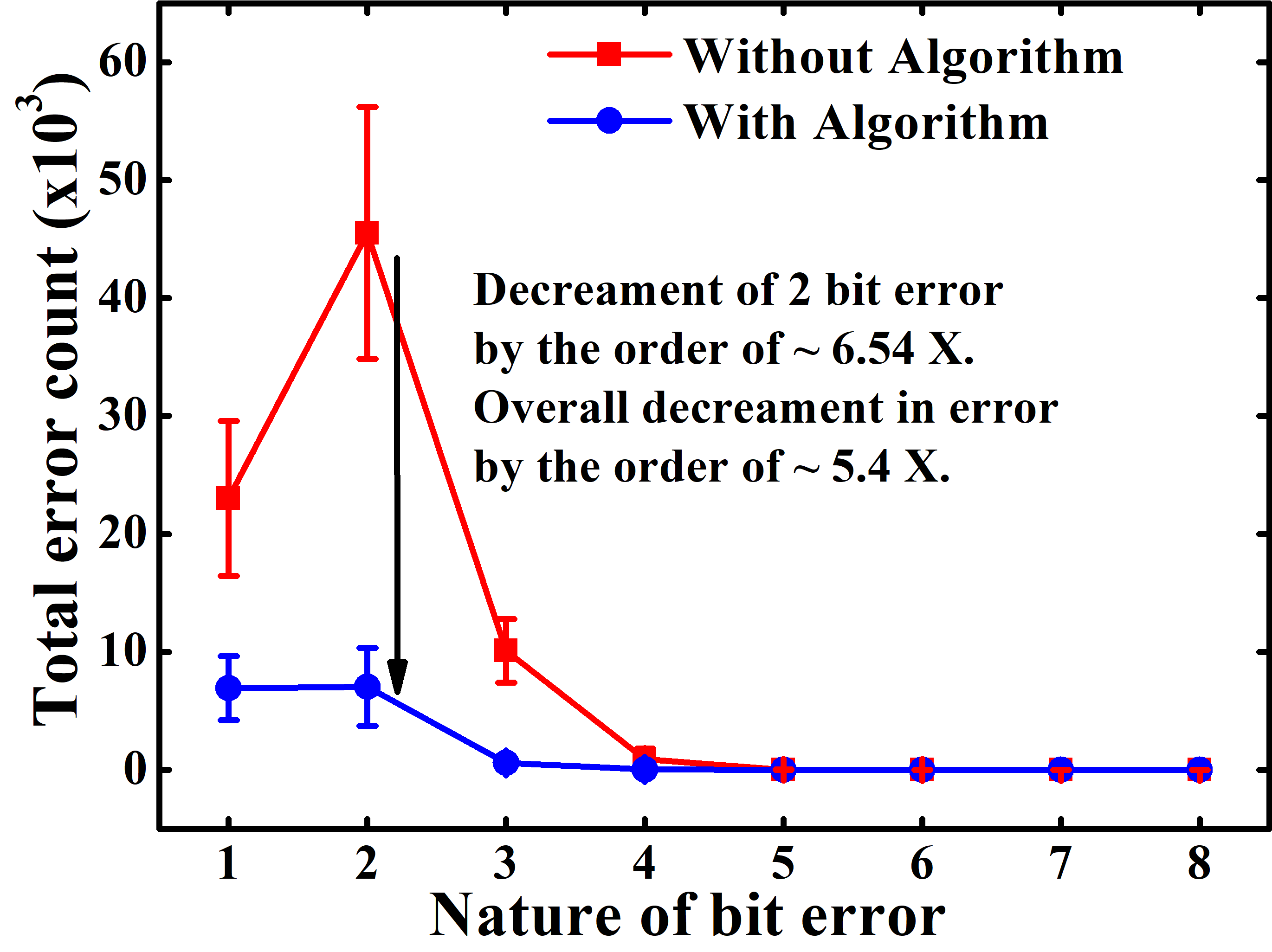}
    \caption{Comparison of decrease in nature of bit errors by implementing algorithms (soft technique) for error reduction.}
    \label{error_plot}  
    \vspace{-5mm}
\end{figure}

\subsection{Current variation with aging}
Page write current variation due to aging effect in CBRAM NVM chips is shown in Fig. \ref{aging_current_cbram}(a). It is observed that current consumption increases with aging. One of the probable reasons that describe the phenomenon of increase in page write current consumption over cycling is the write-verify-write (WvW) scheme for write operation. In WvW scheme step voltages with increase in pulse width and voltage amplitude are applied to increase the success rate of write operation \cite{chen2016128kb}. WvW scheme also increases the write latency. Significant results have been observed for CBRAM technology. In order to observe the WvW technique, write latency is also measured (as shown in inset Fig. \ref{aging_current_cbram} (b)). The increasing trend of write latency values over cycling supports the effect of aging. However, it should be noted here that no significant change in current consumption over cycling is observed for the other NVM technologies (Toggle MRAM, FeRAM, and ReRAM).

\subsection{Endurance Characterization}
The data-sheet specified endurance of CBRAM chip is 100k write cycles while the minimum write endurance for other NVM technologies used for the study is 1.2$\times$ $10^6$ cycles. We focused our study on analyzing the nature and distribution of bit errors that occur in NVM technologies due to over-stressing a particular location. We select CBRAM chips for error characterization at page and byte level granularity for ou study.
Fig. \ref{error_colour_page} (a) and (b), shows the distribution of nature of errors occurred in a page and a byte level granularity respectively for random data write operations performed over 200k cycles (2X data-sheet specifications). It can be observed that the distribution of error in a page is random. The total error count for a particular byte in a page varies randomly. However, the total count for 2-bit error in a page is more compared to other types of bit errors. Moreover, higher bit errors (3-bit, 4-bit, etc.) counts are few. The probable reason for this specific nature of bit error is due to the implementation of specific bit error ECC within the chip. We have implemented Flip-N-Write (FNW) algorithm to analyze the effect of soft techniques in reducing the nature of error. It is observed from Fig. \ref{error_plot} that the implementation of FNW decreases the 2-bit error by $\sim$ 6.54X and overall total error count by $\sim$ 5.4X. Thus, based on the nature of error and the implementation of soft techniques, the endurance of the NVM chips can be increased significantly \cite{bhattacharya2017advanced}. This analysis helps in designing the soft/hard techniques like ECC within/next to the controller to enhance the endurance of the emerging NVM chips.

\section{Case Study: FPGA based NN application}
\subsection{Methodology}
\begin{figure}
  \centering
  \includegraphics[scale=0.25]{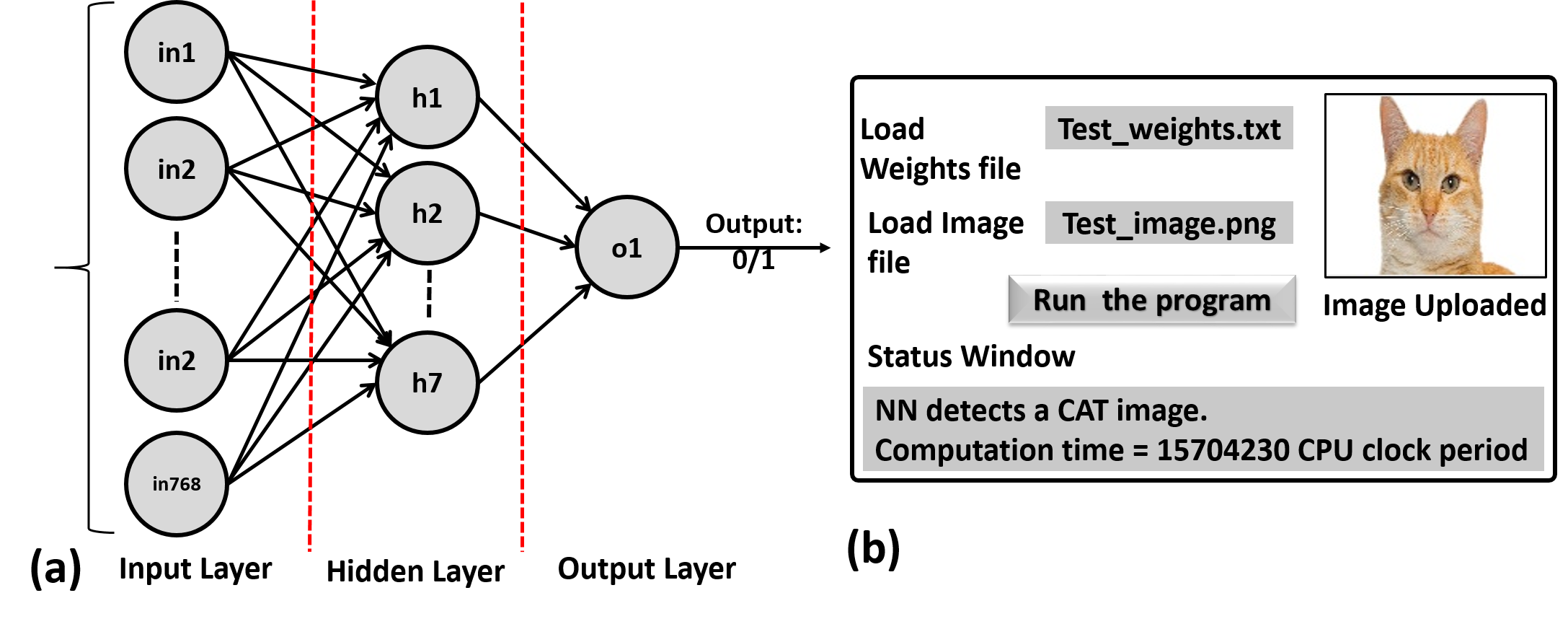}
    \caption {(a) NN Model implemented on FPGA platform for characterizing different NVM technologies. (b) Software GUI for uploading weights and image for AI application.}
    \label{nn_model}
\end{figure}

In this section, we benchmarked the emerging NVM chips for NN application using the proposed FPGA based test platform. We focused on characterizing the weights' write latency and current consumption of the NVM chips for AI applications. The NN model used for the study consists of three layers: (i) input layer, (ii) hidden layer, and (iii) output layer (Fig. \ref{nn_model} (a)). An RGB image (16$\times$16$\times$3 byte) is taken as an input. Therefore, the input layer consists of 768 neurons, where each neuron corresponds to a particular byte of the input image. 
The hidden layer is composed of 7 neurons feeding to a single output neuron. The neuron output from the hidden layer to the output layer is described by: 
\begin{equation}
y =g{\mathlarger{\sum}}_{j=1}^{n} a_j\times w_j + b
\label{eq1}
  \end{equation}
where $n$ is the number of nodes in the previous layer, $a_j$ is the activation output represented as an 8-bit unsigned integer, $w_j$ and $b$ is the weight, and bias respectively and are represented by 8-bit signed integer. The activation function ($g(x)$) is a step function described by:
\begin{equation}
    g(x)=\left\{   
    \begin{array}{@{}ll@{}}
    0, & x<0 \\
    255,& x\geq0
    \end{array}\right.
    \label{eq2}
\end{equation}

The network is trained using differential evolution (DE) algorithm \cite{ilonen2003differential} for the dataset used in \cite{coursera}. The algorithm is suitable for network having integer weights, biases and non-linear activation function \cite{plagianakos1999neural}. The integer weights and biases allow easier hardware implementation in FPGA.
\subsection{Hardware Implementation}
\begin{figure}
  \centering
  \includegraphics[scale=0.33]{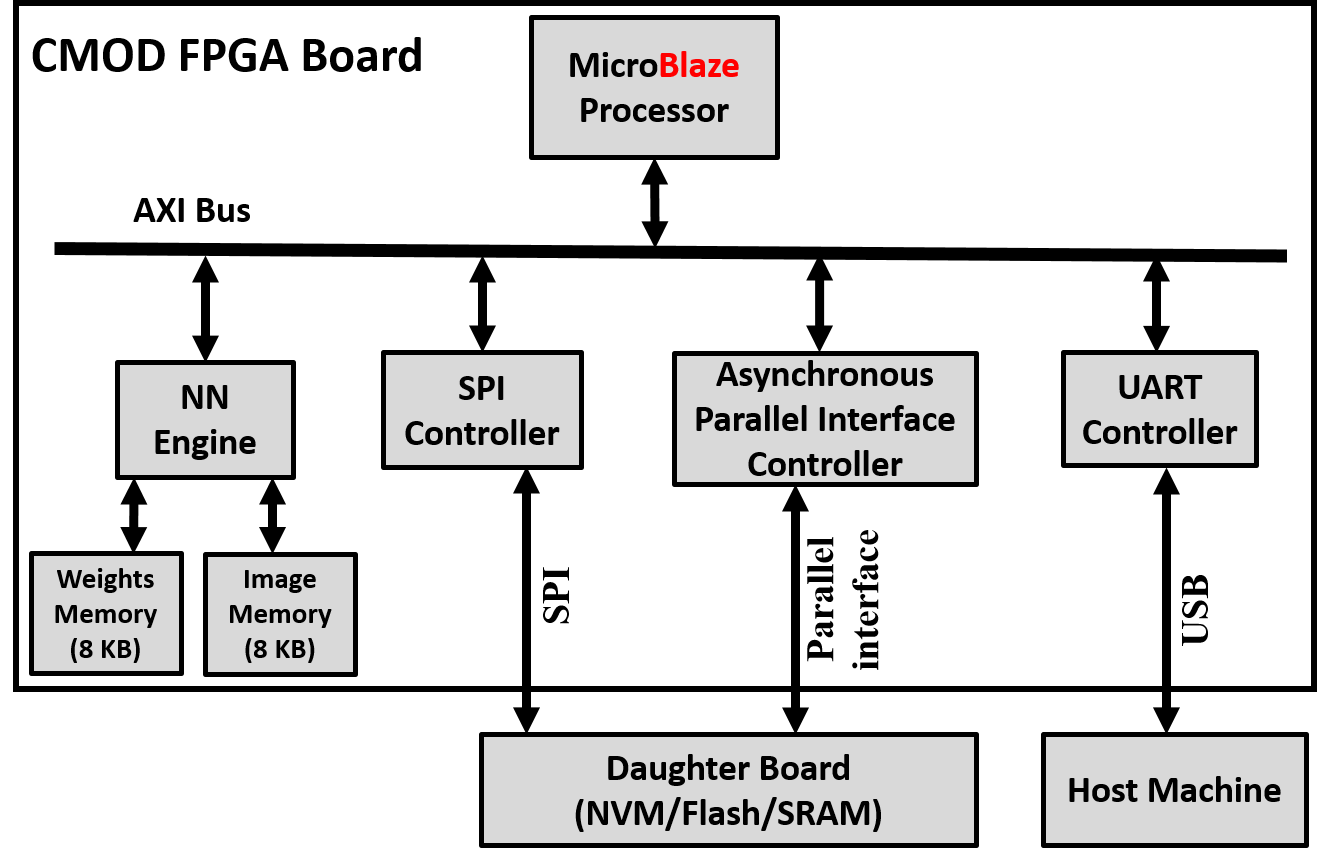}
    \caption{Block diagram of FPGA based setup for NN application using emerging NVM.}
    \label{nn_setup}
     \vspace{-5mm}
\end{figure}
\begin{table}[!t]
\renewcommand{\arraystretch}{1.3}
\caption{Latency and current comparisons of emerging NVM chips with Flash and SRAM technologies for NN application.}
\begin{threeparttable}[!t]
\label{table2}
\begin{tabular}{|c|c|c|c|c|}
\hline
\textbf{Memory Type} & \multicolumn{2}{c|}{\textbf{Latency (a.u.)}} & \multicolumn{2}{c|}{\textbf{Current (a.u.)}} \\ 
 \cline{2-5}
& Weights  & NN &Average  & Average Byte \\
& Write& Application \tnote a&Erase& Write\\
\hline
Toggle MRAM& 5.2x10$^{-5}$ & 0.05& 0 & 1\\
\hline
FeRAM & 0.02 & 0.82& 0 & 0.02\\
\hline
CBRAM & 0.07 & 0.82& 0 & 0.03\\
\hline
ReRAM & 1 & 0.99& 0 & 0.04\\
\hline
Flash & 0.26 \tnote b  & 1 \tnote b & 1 &1\\
\hline
SRAM & 5.2x10$^{-5}$ & 0.05 & 0 & 1\\
\hline
\end{tabular}
\begin{tablenotes}
    \item[a] Latency value includes read latency from emerging NVM chips to BRAM and NN computational latency.
     \item[b] Latency value includes summation of erase and write operation together.
     \end{tablenotes}
    \end{threeparttable}
       \vspace{-5mm}
\end{table}
The hardware architecture, as shown in Fig. \ref{nn_setup}, is implemented on CMOD FPGA \cite{CMOD} evaluation board. The main difference between the FPGA based NN hardware architecture and the experimental setup used for chip characterization is the addition of NN engine with the former architecture. NN engine is a generic engine that can be used to implement multi-layer NN to predict the test image uploaded by the GUI based software (Fig. \ref{nn_model} (b)). For our application, we implement a 2-layer NN. Two internal block RAMs (BRAM) of the FPGA are used to interact with the NN engine. We term the BRAMs as image memory and weight memory (Fig. \ref{nn_setup}). Image memory contains the scaled input RGB image. Weight memory contains the trained weights and biases of each layer of fully connected NN read from the NVM chip. Following operations are performed while prediction of the test image (i) Initially, the trained weights and the biases are uploaded in the emerging NVM chips. (ii) The pixel values of the input RGB image are uploaded in the image memory.  (iii) NN engine is started for image prediction. The NN engine on initiation performs inference by reading the trained weights and biases and loads it to the FPGA BRAM (weight memory) for faster execution. Finally, it starts calculating the results sequentially of each hidden neuron followed by an output neuron to distinguish desired images from undesired images.
\subsection{Experimental Results}
Extensive experiments are performed to calculate weights' write latency and current consumption of the NVM technologies while performing NN applications.
Comparative study of the NVM technologies with the state-of-art Flash technology and SRAM is also performed. The DE algorithm used for the application provides approximately 72.25 \% accuracy on the data set containing 209 images. During the initialization phase, number of weight/bias byte write operations to NVM chip is 5391 and number of image byte write operations on FPGA BRAM is 768. Similar number of read operations are performed after initiating the NN engine for test image prediction. Table \ref{table2} illustrates the study of the write latency and the electric current consumption using the proposed setup for different memory technologies. It can be observed that off-the-shelf emerging NVM chips are proficient candidates for replacement of Flash and SRAM technologies.

\section{Conclusion}
The paper presents a unified FPGA based test platform for characterizing different off-the-shelf emerging NVM technologies. Detailed electrical characterization and benchmarking study for multiple NVM chips using the test setup is performed. Current consumption due to different data patterns and aging effect is analyzed on emerging NVM technologies such as MRAM, FeRAM, ReRAM, and CBRAM. Moreover, nature of error and the distribution of error are analyzed at byte and page level granularity. Finally, the proposed test platform is utilized for NN image classification application. A comparative study of the emerging NVM chips with state-of-art Flash and SRAM technology is performed. Obtained results show that off-the-shelf emerging NVM chips are suitable candidates for future memory applications.
\section*{ACKNOWLEDGMENT}
This work was supported in part by MHRD, SERB-CRG/2018/001901,
and CYRAN.
\bibliographystyle{IEEEtran}
\bibliography{IEEEabrv,main.bib}

\end{document}